\documentclass[prl,superscriptaddress,twocolumn,floatfix]{revtex4}

\usepackage{graphicx}
\usepackage{psfrag}
\usepackage{amsmath}

\begin{document}
\newtheorem{theorem}{Theorem}
\newtheorem{acknowledgement}[theorem]{Acknowledgment}
\newtheorem{algorithm}[theorem]{Algorithm}
\newtheorem{axiom}[theorem]{Axiom}
\newtheorem{claim}[theorem]{Claim}
\newtheorem{conclusion}[theorem]{Conclusion}
\newtheorem{condition}[theorem]{Condition}
\newtheorem{conjecture}[theorem]{Conjecture}
\newtheorem{corollary}[theorem]{Corollary}
\newtheorem{criterion}[theorem]{Criterion}
\newtheorem{definition}[theorem]{Definition}
\newtheorem{example}[theorem]{Example}
\newtheorem{exercise}[theorem]{Exercise}
\newtheorem{lemma}[theorem]{Lemma}
\newtheorem{notation}[theorem]{Notation}
\newtheorem{problem}[theorem]{Problem}
\newtheorem{proposition}[theorem]{Proposition}
\newtheorem{remark}[theorem]{Remark}
\newtheorem{solution}[theorem]{Solution}
\newtheorem{summary}[theorem]{Summary}    
\def\r{{\bf{r}}}
\def\i{{\bf{i}}}
\def\j{{\bf{j}}}
\def\m{{\bf{m}}}
\def\k{{\bf{k}}}
\def\h{{\bf{h}}}
\def\kt{{\tilde{\k}}}
\def\qt{{\tilde{\q}}}
\def\mt{{\hat{t}}}
\def\mG{{\hat{G}}}
\def\mg{{\hat{g}}}
\def\mGa{{\hat{\Gamma}}}
\def\mS{{\hat{\Sigma}}}
\def\mT{{\hat{T}}}
\def\K{{\bf{K}}}
\def\P{{\bf{P}}}
\def\q{{\bf{q}}}
\def\Q{{\bf{Q}}}
\def\p{{\bf{p}}}
\def\x{{\bf{x}}}
\def\X{{\bf{X}}}
\def\Y{{\bf{Y}}}
\def\F{{\bf{F}}}
\def\R{{\bf{R}}}
\def\G{{\bf{G}}}
\def\bG{{\bar{G}}}
\def\mbG{{\hat{\bar{G}}}}
\def\M{{\bf{M}}}
\def\V{\cal V}
\def\tchi{\tilde{\chi}}
\def\tx{\tilde{\bf{x}}}
\def\tk{\tilde{\bf{k}}}
\def\tK{\tilde{\bf{K}}}
\def\tq{\tilde{\bf{q}}}
\def\tQ{\tilde{\bf{Q}}}
\def\si{\sigma}
\def\ep{\epsilon}
\def\hep{{\hat{\epsilon}}}
\def\al{\alpha}
\def\be{\beta}
\def\ep{\epsilon}
\def\bep{\bar{\epsilon}_\K}
\def\mep{\hat{\epsilon}}
\def\up{\uparrow}
\def\de{\delta}
\def\De{\Delta}
\def\up{\uparrow}
\def\dwn{\downarrow}
\def\ksi{\xi}
\def\etha{\eta}
\def\product{\prod}
\def\goto{\rightarrow}
\def\switch{\leftrightarrow}

\title{Efficient calculation of the antiferromagnetic phase diagram of
the 3D Hubbard model} 

\author{P. R. C. Kent} \affiliation{University of Tennessee, Knoxville,
 TN 37996}
\author{M.~Jarrell} \affiliation{Department of Physics, University of
  Cincinnati, Cincinnati, OH 45221}
\author{T. A. Maier}\affiliation{Computer Science and Mathematics
  Division, Oak Ridge National Laboratory, Oak Ridge, TN 37831}
\author{Th.~Pruschke}\affiliation{Institut f\"ur Theoretische Physik,
  Universit\"at G\"ottingen, Friedrich-Hund-Platz 1, 37077
  G\"ottingen, Germany}

\date{\today}

\begin{abstract} 
  The Dynamical Cluster Approximation with Betts clusters is used to
  calculate the antiferromagnetic phase diagram of the 3D Hubbard
  model at half filling.  Betts clusters are a set of periodic clusters
  which best reflect the properties of the lattice in the thermodynamic limit and
  provide an optimal finite-size scaling as a function of cluster
  size.  Using a systematic finite-size scaling as a function of
  cluster space-time dimensions, we calculate the antiferromagnetic
  phase diagram.  Our results are qualitatively consistent with the
  results of Staudt et al. [Eur. Phys. J. B 17 411 (2000)], but require
  the use of much smaller clusters: 48 compared to 1000.
\end{abstract}

\maketitle

The accurate and efficient solution of lattice Hamiltonians such as
the Hubbard model is a long standing challenge in the theoretical
condensed matter community. These lattice models are routinely solved
on a finite periodic lattice, for example with Monte Carlo, and the
calculated properties extrapolated to the infinite limit. Due to the
numerical expense in solving these models for large lattices, it is
imperative to choose lattices that are efficient for the estimation
and extrapolation of the physical properties of interest.


In this paper we use the Dynamical Cluster Approximation (DCA)
\cite{hettler:dca1,hettler:dca2,maier:dca1} (for a review see 
\cite{maier:rev}) to explore the antiferromagnetic instability in the 3D 
Hubbard model at half filling, with
\begin{equation}
  \label{eq:HM}
  H=-t \sum_{\langle ij\rangle, \sigma} c^\dagger_{i\sigma}c^{\phantom\dagger}_{j\sigma} +U\sum_i
  (n_{i\uparrow}-\frac{1}{2}) (n_{i\downarrow}-\frac{1}{2})\,,
\end{equation}
where $c^{(\dagger)}_{i\sigma}$ (creates) annihilates an electron with
spin $\sigma$ on site $i$, $n_{i\sigma}$ is the corresponding number
operator, $t$ the hopping amplitude between nearest neighbors $\langle
i,j\rangle$
and $U$ the on-site Coulomb repulsion.  We solve this
model on a series of finite clusters chosen according to the
criteria proposed by Betts et al.\cite{dbetts:2d,dbetts:3d}. We obtain
converged results extrapolating from clusters of up to only 48 sites, which
are in agreement with the calculations of Staudt et
al.\cite{RStaudtEPJB2000}, who used conventional cubic lattices of up
to 1000 sites and obtained the N\'eel temperature via the specific
heat.

To solve the Hamiltonian (\ref{eq:HM}) we utilized the DCA
\cite{maier:rev}. For a $3D$ system the DCA maps the original lattice
model onto a periodic cluster of size $N_c=L_c^3$ embedded in a
self-consistent host. Thus, correlations up to a range $\xi\lesssim
L_c$ are treated directly, while the longer length scale physics is
described at the mean-field level.  With increasing cluster size, the
DCA systematically interpolates between the single-site dynamical
mean-field result and the exact result, while remaining in the
thermodynamic limit. We solve the cluster problem using Quantum Monte
Carlo (QMC) \cite{jarrell:dca3}. At half-filling there is no QMC sign
problem; the only systematic error in the Monte Carlo is the time step
error, which can be extrapolated away.

In order to calculate the phase diagram of the system in the thermodynamic 
limit, we employ the scaling ansatz $\xi(T_{\rm N}^{\rm DCA})=L_c$ where $T_{\rm N}^{\rm DCA}$ is the
N\'eel temperature obtained from a DCA calculation with a cluster of linear cluster size $L_c$.  This
form is justified if we envision the lattice as perfectly tiled by a periodic array of non-overlapping
clusters.  This system becomes ordered when the antiferromagnetic correlations of the cluster reach the
linear cluster size. According to this ansatz  
$\xi(T_{\rm N}^{\rm DCA})\propto |T_{\rm N}^{\rm DCA}-T_{\rm N}|^{-\nu}\propto L_c$, so that
\begin{equation}\label{eq:scaling}
        T_{\rm N}^{\rm DCA} = T_{\rm N} + BN_c^{-1/3\nu}
\end{equation}
where  $T_{\rm N}$ is the true
antiferromagnetic transition temperature in the thermodynamic limit. The 
exponent is well-known for the 3D Heisenberg model, where one finds
$\nu\approx 0.71$\cite{asandvik:3dheis}.

Betts et al.  \cite{dbetts:2d,dbetts:3d} systematically studied the 2D
and 3D Heisenberg models on finite size clusters and developed a
grading scheme to determine which clusters should be used in finite
size simulations.  The main qualification is the ``perfection'' of the
near-neighbor shells: a measure of the completeness of each neighbor
shell compared to the infinite lattice.  A perfect finite size cluster
has all neighbor shells up to the k-th shell complete, the k-th shell 
is incomplete, and all shells k+1 and higher are empty.  The absolute
deviation from this criteria is defined as the imperfection.  I.e. if 
the cluster neighbor configuration is as described, except that
the k-1 shell is missing one entry, the cluster imperfection is one.
The second qualification is the cubicity\cite{dbetts:3d}, 
$C=max(c_1,c_1^{-1}) max(c_2,c_2^{-1})$, where 
$c_1=3^{1/2}l/d$ and $c_2=2^{1/2}l/f$ are
defined by the geometric mean
of the lengths of the four body diagonals of the cluster, 
$d=\left(d_1d_2d_3d_4\right)^{1/4}$, the six face diagonals
$f=(f_1f_2f_3f_4f_5f_6)^{1/6}$, and the edges $l=(l_1l_2l_3)^{1/3}$.
$C=1$ for a cube, so the deviation of $C$ from one is a measure of the
cubic imperfection.  In finite size scaling
calculations of the order parameter and ground state energy, they
found that the results for the most perfect clusters fall on a scaling
curve, while the imperfect clusters generally produce results off the
curve. We generated additional 3D clusters following these guidelines
for clusters larger than the 27 site clusters previously
published\cite{dbetts:3d}, from which we adopt the labeling
conventions and cluster geometries. In Table \ref{tab:clus3d} we list
clusters up to 70 sites, their perfection and cubicity. In each case,
we chose either the bipartite (labeled B) or non-bipartite 
cluster (labeled A) with the smallest imperfection and 
cubicity closest to one, in this order of priority. For example, the 38 
site cluster 38B is bipartite, perfect, consisting of only complete neighbor 
shells, and has a cubicity of 1.087. Since we are interested in a 
calculation of $T_{\rm N}^{\rm  DCA}$, we utilized only bipartite 
clusters in the present calculations.

\begin{table}[htbp]
  \begin{tabular}{cccccc}
$N_c$ & $\vec{a}_1$ & $\vec{a}_2$ & $\vec{a}_3$ & Imperfection & Cubicity \\ \hline \hline
28A&( 1, 1, 3)&( 3,-1, 1)&( 1, 2,-2)&  0&1.063  \\
28B&( 1, 1, 2)&( 3, 2,-1)&( 1,-3, 2)&  5&1.018  \\
30A&( 1, 2, 2)&( 2, 2,-2)&( 2,-2, 1)&  0&1.007  \\
30B&( 1, 1, 2)&( 3, 1,-2)&( 3,-2, 1)&  4&1.012  \\
32A&( 1, 1, 3)&( 2, 2,-2)&( 2,-2, 1)&  0&1.022  \\
32B&( 1, 2, 3)&( 2, 0,-2)&( 2,-2, 2)&  3&1.028  \\
34A&( 1, 1, 3)&( 3,-2, 0)&( 1, 2,-2)&  0&1.009  \\
34B&( 1, 0, 3)&( 2, 2,-2)&( 1,-3,-2)&  2&1.057  \\
36A&( 1, 2, 2)&( 3, 0,-2)&( 2,-2, 2)&  0&1.004  \\
36B&( 1, 0, 3)&( 3, 2,-1)&( 2,-2,-2)&  3&1.040  \\
38A&( 1, 1, 3)&( 3, 1,-3)&( 2,-2, 1)&  0&1.002  \\
38B&( 1, 2, 3)&( 3,-1,-2)&( 2,-2, 2)&  0&1.087  \\
40A&( 1, 2, 2)&( 3, 1,-2)&( 1,-3, 2)&  0&1.003  \\
40B&( 1, 2, 3)&( 2, 2,-2)&( 2,-2, 2)&  3&1.041  \\
42A&( 1, 2, 2)&( 3, 0,-2)&( 0, 3,-3)&  0&1.005  \\
42B&( 1, 2, 3)&( 3,-1, 2)&( 2, 2,-2)&  2&1.056  \\
44A&( 1, 2, 2)&( 3, 2,-2)&( 3,-2, 1)&  0&1.010  \\
44B&( 1, 2, 3)&( 3, 2,-1)&( 2,-2, 2)&  3&1.035  \\
46A&( 1, 1, 3)&( 3, 2,-2)&( 3,-2, 0)&  0&1.014  \\
46B&( 1, 2, 3)&( 3, 1,-2)&( 2,-2, 2)&  4&1.017  \\
48A&( 1, 1, 3)&( 3, 2,-2)&( 2,-3,-1)&  0&1.009  \\
48B&( 1, 2, 3)&( 3,-2, 1)&( 2, 2,-2)&  5&1.002  \\
50A&( 1, 1, 3)&( 3, 2,-2)&( 2,-3, 1)&  1&1.005  \\
50B&( 1, 2, 3)&( 3, 2,-1)&( 2,-3, 1)&  6&1.018  \\
52A&( 2, 2, 3)&( 3, 2,-2)&( 3,-2,-2)&  1&1.109  \\
52B&( 1, 2, 3)&( 3, 1,-2)&( 2,-3, 1)&  7&1.003  \\
54A&( 1, 2, 3)&( 3, 0,-3)&( 3,-2, 2)&  2&1.063  \\
54B&( 1, 2, 3)&( 3,-3, 0)&( 2, 2,-2)&  8&1.005  \\
56A&( 1, 1, 3)&( 3, 2,-2)&( 3,-3,-1)&  3&1.003  \\
56B&( 1, 2, 3)&( 3, 2,-3)&( 3,-1, 2)&  9&1.029  \\
58A&( 1, 1, 3)&( 3, 2,-2)&( 3,-3, 1)&  3&1.014  \\
58B&( 1, 2, 3)&( 3,-3, 2)&( 2, 2,-2)& 10&1.011  \\
60A&( 2, 0, 3)&( 2, 3,-2)&( 2,-3,-2)&  4&1.001  \\
60B&( 1, 2, 3)&( 3,-3, 2)&( 2, 1,-3)& 11&1.011  \\
62A&( 1, 3, 3)&( 3, 2,-2)&( 3,-3,-1)&  5&1.087  \\
62B&( 1, 2, 3)&( 3, 2,-1)&( 3,-3, 2)& 12&1.003  \\
64A&( 1, 2, 3)&( 3, 1,-3)&( 2,-3, 2)&  6&1.013  \\
64B&( 1, 2, 3)&( 3, 2,-3)&( 2,-3, 1)& 12&1.010  \\
66A&( 1, 3, 3)&( 3, 3,-1)&( 2,-3, 2)&  5&1.067  \\
66B&( 1, 2, 3)&( 3, 0,-3)&( 3,-3, 2)& 11&1.026  \\
68A&( 1, 3, 3)&( 3, 3,-1)&( 3,-2, 2)&  4&1.055  \\
68B&( 1, 2, 3)&( 3, 3,-2)&( 2,-3, 3)& 10&1.054  \\
70A&( 1, 3, 3)&( 3, 3,-2)&(-2, 2,-3)&  3&1.063  \\
70B&( 1, 2, 3)&( 3, 3,-2)&( 3,-2, 3)&  9&1.034  \\
  \end{tabular}
  \caption{3D cluster geometries, imperfection and cubicity of the 
  best non-bipartite (A clusters) and bipartite (B clusters). The
  $a_{i}$ denote the cluster lattice vectors.} 
  \label{tab:clus3d}
\end{table}

\begin{figure}[htbp]
  \centering
  \includegraphics*[width=3.5in]{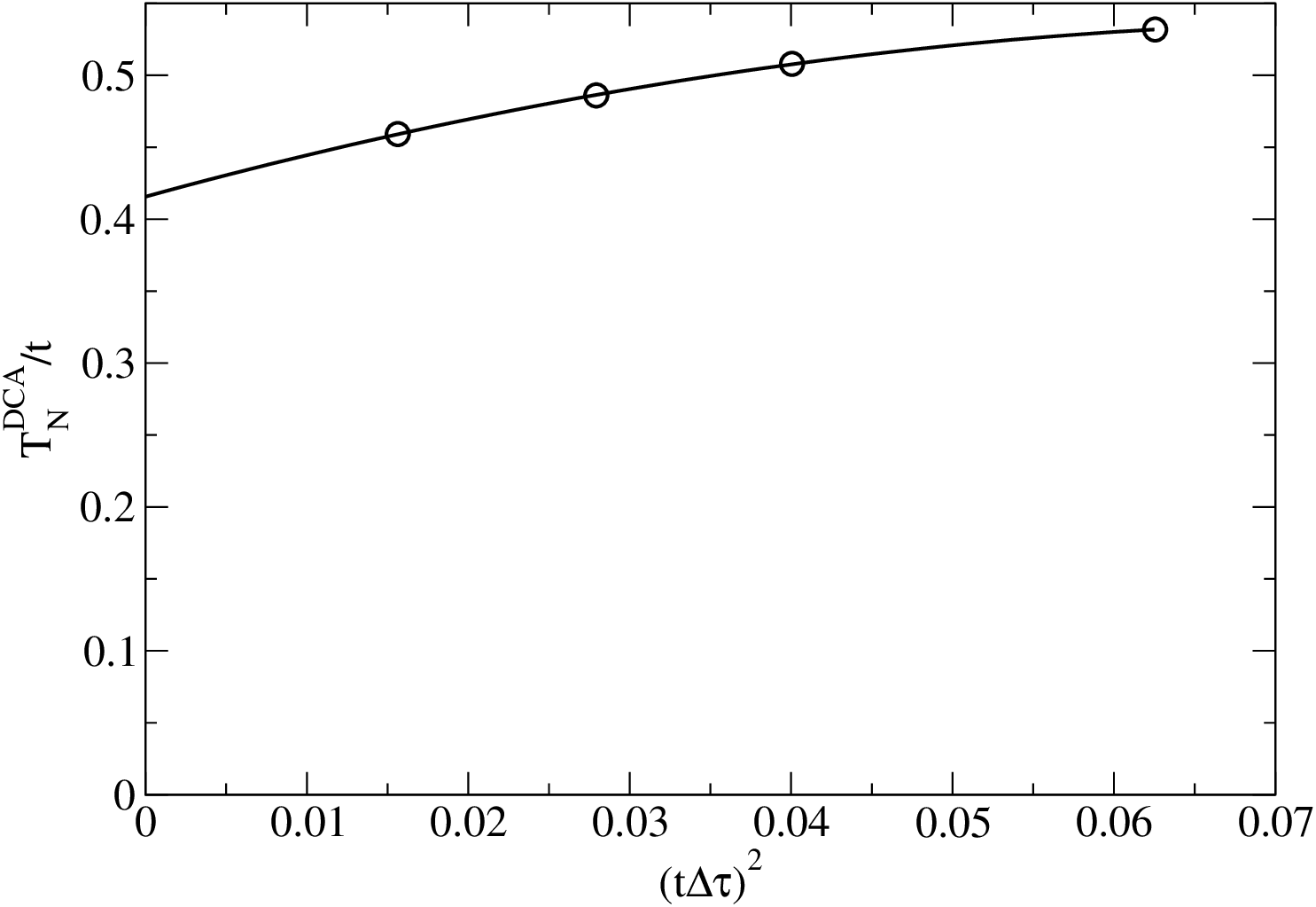}
  \caption{$T_N$ versus $\Delta\tau^2$ when $U/t=8$ for cluster 18A}
  \label{fig:dtau}
\end{figure}

To obtain the antiferromagnetic phase diagram we performed a series of
DCA calculations as a function of U/t, cluster size, and Monte-Carlo
time step $\Delta\tau$.
For a given U/t and cluster size, we calculated $T_{\rm N}^{\rm DCA}$ by finding the
divergence of the staggered susceptibility as a function of $\Delta\tau$, and
extrapolated the value obtained to $\Delta\tau=0$.
As an example, we show in Fig.\ref{fig:dtau} the N\'eel temperature
$T_{\rm N}^{\rm DCA}(\Delta\tau)$
for an 18 site cluster for $U/t=8$. One finds a
significant $\Delta\tau$ dependence which makes an extrapolation to
$\Delta\tau=0$ mandatory.

\begin{figure}[htbp]
\centering 
\includegraphics*[width=3.5in]{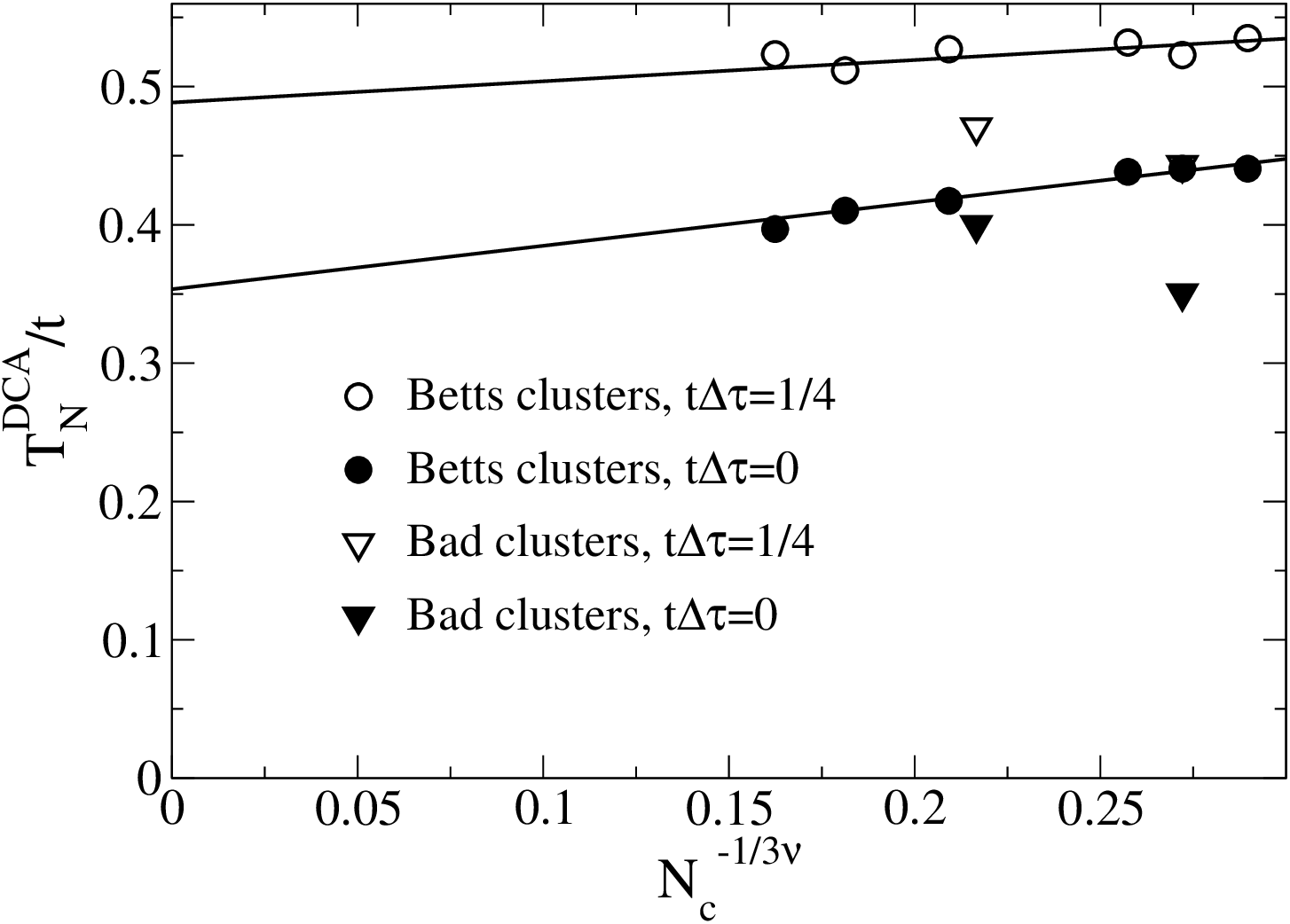} 
\caption{Cluster size scaling of $T_N$ when $U/t=8$ and $t\Delta\tau=1/4$ 
(open circles) and the result extrapolated to $t\Delta\tau=0$ (full circles) 
as in Fig.~\ref{fig:dtau}.} 
\label{fig:Tc}
\end{figure}

Performing this extrapolation for the series of bipartite clusters
from Table \ref{tab:clus3d} for $U/t=8$, we obtain the values for
$T_{\rm N}^{\rm DCA}$ collected in Fig.~\ref{fig:Tc} (full circles). For
comparison we also included the results for a finite
$t\cdot\Delta\tau=1/4$ (open circles). This unextrapolated data
actually lies above the Heisenberg result of T/t=0.48. One clearly
sees that a proper scaling to $\Delta\tau=0$ is necessary to obtain
both the correct qualitative {\em and} quantitative behavior of
$T_{\rm N}^{\rm DCA}(N_c)$. The full curves in Fig.~\ref{fig:Tc} were
obtained with the scaling ansatz (\ref{eq:scaling}) using the $\nu$
for the 3D Heisenberg model.  It yields a linear scaling curve within
our error bars.

To assess the value of Betts clusters, we also study two bad
clusters, 16Z and 26Z, identified in Table \ref{tab:clus3dbad}.
\begin{table}[htbp]
  \begin{tabular}{cccccc}
$N_c$ & $\vec{a}_1$ & $\vec{a}_2$ & $\vec{a}_3$ & Imperfection &
Cubicity \\ \hline \hline
16Z&( 2, 0, 0)&( 0, 2, 0)&( 0, 0, 4)&  7& 1.209  \\
26Z&( 1, 2, 3)&( 3, 3,-2)&( 3,-2, 3)& 14& 1.295 \\
  \end{tabular}
  \caption{3D cluster geometries, imperfection and cubicity of two 
  poor quality bipartite clusters.}
  \label{tab:clus3dbad}
\end{table}
Although these clusters are bipartite, they are highly imperfect.
Both are missing independent neighbors in the first shell (each have
4; whereas a complete first shell has 6 neighbors).  As a result of the
periodic boundary conditions on the cluster, this causes the near-neighbor
fluctuations to be over estimated.  As a result, the estimates of $T_{\rm N}$ 
from these clusters, shown Fig.~\ref{fig:Tc} for a finite 
$t\cdot\Delta\tau=1/4$ (open triangles) and for the data extrapolated to 
$\Delta\tau=0$ (filled triangles), fall well below the scaling curve 
established by the best cluster geometries listed in table \ref{tab:clus3d}.
In general, in this and in other calculations, we find that the less perfect
clusters tend to overestimate the effects of fluctuations.

\begin{figure}[htbp]
  \includegraphics*[width=3.5in]{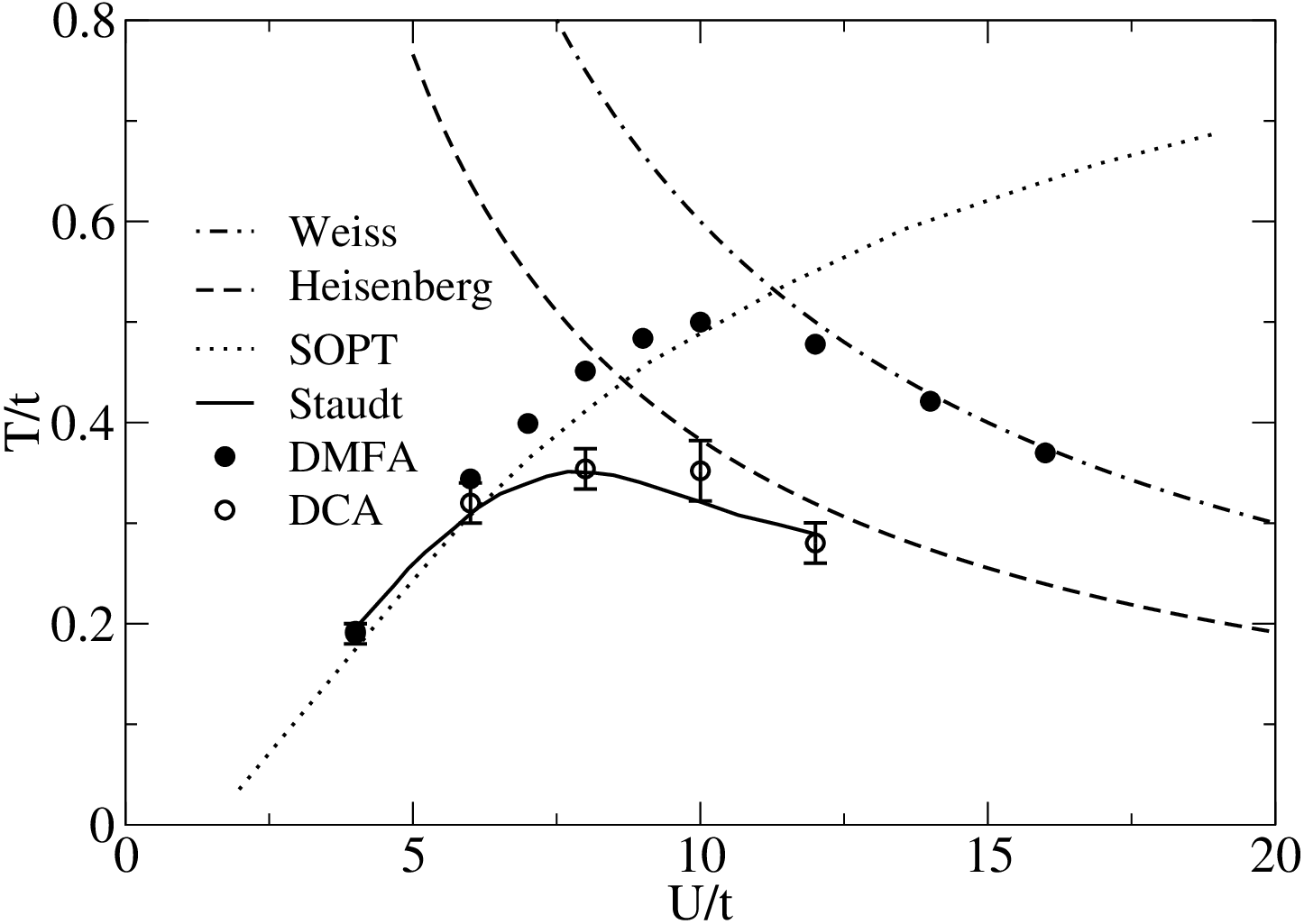}  
  \caption{Antiferromagnetic phase diagram of the 3D Hubbard model from our
  results and different approximations.  }
  \label{fig:phasediagram}
\end{figure}

Finally, Fig.\ref{fig:phasediagram} displays the calculated
antiferromagnetic phase diagram obtained from the DCA and extrapolated
to $\Delta\tau=0$ and $N_c=\infty$ (open circles with error bars). For
comparison, we included results from other methods: The dynamical
mean-field approximation (DMFA, full circles), Staudt et
al.\cite{RStaudtEPJB2000} (full curve), second order perturbation
theory (SOPT, dotted curve)\cite{niki:SOPT_3DHM,pvd:SOPT_3DHM}, the 
Heisenberg model (dashed curve)\cite{GSRushbrookeInPhaseTransitions1974} 
and the Weiss mean-field theory for the Heisenberg model (dash-dotted 
curve).  We took $J=4t^2/U$ for both Heisenberg calculations.  The 
results from Staudt et al.\ are reproduced with good accuracy, but with 
much smaller clusters. The DMFA result is obtained through the methods
described above when $N_c=1$.  Both the DMFA and the Weiss mean field
are local approximations which neglect the effect of non-local
fluctuations.  As expected, they agree in the strong coupling regime,
$U>12t=W$ ($W$ is the bandwidth).  Both DMFA and SOPT are only
accurate at small $U/t$, indicating that non-local fluctuations are
not important for small $U$. At large $U/t$ the DCA results for
$T_{\rm N}$ approach the curve for the Heisenberg model, as expected.
However, for intermediate and large values of $U/W$, the deviation
between the present results and the mean-field results is as large as
$30\%$ or more, indicating that the effects of non-local fluctuations
are significant.

In conclusion, we have calculated the antiferromagnetic phase diagram
of the 3D Hubbard model at half filling using the Dynamic Cluster
Approximation and Betts clusters. Well converged results are found for
relatively small cluster sizes due to the well-chosen geometries of
these clusters. The dramatically increased efficiency of these
clusters compared to typically used cluster geometries, such as cubic
lattices, suggests that these clusters should be more widely used for
lattice calculations.

We acknowledge useful discussions with J.P.\ Hague and G.\ Stewart.  This 
research used computational resources of the Center for Computational 
Sciences,
and was sponsored by the offices of Basic Energy Sciences and Advanced 
Scientific Computing Research, U.S. Department of Energy. Oak Ridge 
National Laboratory is managed by UT-Battelle, LLC under Contract No.  
DE-AC0500OR22725. This research was supported by the NSF Grant No.  
DMR-0312680 and supported in part by NSF cooperative agreement 
SCI-9619020 through resources provided by the San Diego Supercomputer 
Center.


\end{document}